\documentstyle[prl,aps,multicol]{revtex}
\begin{document}
\tightenlines

\title { A New Singularity Theorem in Relativistic Cosmology}


\author { A. K. Raychaudhuri}

\address {Relativity and Cosmology 
Center, Department of Physics, Jadavpur University, Calcutta 700032, India}
\date{today}
\maketitle


\medskip

\begin{abstract}

It is shown that 
if the timelike eigenvector of the Ricci tensor be
hypersurface orthogonal so that  the space time
allows a foliation into space sections then the space average of each of the 
scalar that appear in the Raychaudhuri equation vanishes provided
the strong energy condition holds good. 
This result is presented in the form of a singularity theorem.

\end{abstract}



\begin{multicols} {2}

Quite a number of theorems on the singularity in cosmological solutions exist in the literature.
These are concerned both with the definition of singularity as well as
the condition leading to its occurrence. 
The intuitive definition of singularity 
is an unacceptable behaviour of physical variables like their blowing up or
abrupt discontinuity involving some breakdown of conservation principles.
Of course such peculiarities will be reflected in the geometry of space time. 
However it has been argued that such ``singularities'' may be removed out of 
sight  by introducing coordinate systems whose domain do not
include the "singular regions''. To take care of such situations and
also for mathematical convenience a definition of singularity has emerged which 
identifies singular space times as those in which some null or time like 
geodesic is incomplete. 
Without making a critical discussion on this definition,
we reproduce a formulation of Hawking and Penrose as a standard singularity 
theorem. 
It states \cite{HW}

Space time is not timelike and null geodetically complete if (1)
$R_{\alpha \beta} k^\alpha k^\beta \ge 0$ for every 
non space like vector $k^\alpha$; $R_{\alpha \beta} $ being the Ricci 
tensor (2) Every non space-like geodesic contains a point at which 
$k_{[\alpha }R_{\beta ] \delta \gamma [\rho } K_{\mu ]}K^\gamma k^\delta \ne 0$ where $k^\alpha $
is the tangent vector to the geodesic (3) There are no closed timelike 
curves (4) There exists at least one closed trapped surface. With
the field equations of general relativity, the 
first condition reduces to the strong energy condition (along with an attractive 
gravitation). Although there exists situations like the false vacuum where
the strong energy condition is violated, we shall retain 
condition (1) in our discussion. 
Any violation of condition (3) would mean a breakdown of causality. 
Thus the conditions (1) and (3) may be considered to be 
fundamental elements of standard physics. Not so however are the other two.
Indeed it seems difficult to reconcile the presence of the rather awkward
and complicated condition (2) in the statement of a fundamental theorem. 
Regarding condition(4) we may recall that we usually believe that any realistic model
of the universe should develop a variety of structures at least some of
which would eventually undergo gravitational collapse leading to the formation of trapped
surfaces. To eliminate trapped surfaces from our consideration is to effectively restrict
to structureless universes, unless of course our understanding of stellar evolution 
and gravitational collapse is basically wrong.

In this background came the solution of Senovilla \cite{Seno}. 
The solution is  free of any curvature or physical singularity 
and as shown somewhat later by Chinea et al \cite{chin} is also 
geodetically complete. Of the four conditions in the Hawking
Penrose theorem, only the condition regarding the trapped surfaces did not 
hold good in Senovilla's solution. It thus raised the intriguing question of a more 
useful singularity theorem which will spell out the positive characteristic properties
(physical and/or mathematical) of nonsingular solutions.
An attempt in this direction was by the present author \cite{AKR} where
it was shown that for any nonsingular spatially open non-rotating
universe, the space time averages of each of the scalars that appear
in Raychaudhuri equation must vanish. 

Somewhat later this work was criticised by Saa and Senovilla \cite {SS}
on the ground that for spatially open Friedmann universes with a big bang these
scalars have zero space time averages. This did not contradict the theorem itself 
but merely indicated that the converse is not true. However even this
 criticism can be easily met by 
demanding that the space time average for both the halves of space time- one containing 
future 
 infinity and the other past infinity-must separately vanish.
Senovilla \cite {SS} further made a conjecture that for a still further
restricted class of singularity free cosmological solution, the spatial average of the
energy density shall vanish. However the arguments that he advanced leading to the conjecture
were fallacious. In the present paper we consider that the 
universe is non-rotating in the sense that the timelike eigenvector of 
the Ricci tensor is hypersurface orthogonal 
and give a proof that the spatial averages
of each of the Raychaudhuri scalars indeed vanish for singularity free solution.
Of course for perfect fluids, the timelike eigenvector of the Ricci tensor
coincides with the velocity vector of the fluid which will be
non-rotating because of our assumed condition. For general imperfect
fluids however, there maybe rotation of the matter present even though the eigenvector
of the Ricci tensor is hypersurface  orthogonal. Our present assumption is 
thus somewhat weaker than in \cite{AKR}.

We shall use this hypersurface orthogonal unit timelike eigenvector $v^\alpha$ to set up the
Raychaudhuri equation which now reads, 
\begin{equation}
 {1\over{3}}\theta^2 +2\sigma^2 +\kappa(T_{\alpha \beta}-{1\over{2}}g_{\alpha \beta} T)
v^\alpha v^\beta = -\dot v^{\alpha}_{; \alpha} -\dot \theta   
\end{equation}

The metric with the time coordinate along this vector will have the form 
\begin{equation}
ds^2 = g_{00}dt^2 +g_{ik}dx^idx^k
\end{equation}
The scalars $\theta, \sigma,
\dot v^{\alpha}_{;\alpha}, \dot \theta $ are built up 
from $v^\alpha$ and its covariant derivatives.
Thus with our choice of $v^\alpha$, these scalars will be algebraic 
combination of scalars formed from the Ricci tensor and its covariant 
derivatives.
Hence a blow up of any of these scalars would indicate a blow 
up of some Ricci scalars and hence signal the presence of a singularity. 

We now enunciate and prove the following
theorem:

The space time will be singular in the sense that some scalar built from the 
Ricci tensor will blow up if\\
(a) the strong energy condition is satisfied.\\
(b) the timelike eigenvector of the Ricci tensor is hypersurface
orthogonal. (We are excluding the case of null Ricci tensor.)\\
(c) the space average of any of the scalars occurring in Raychaudhuri
equation does not vanish. 

Here the condition (b) allows a foliation of the space time into space sections and the averages 
referred to in (c) are defined as follows. Space average of any scalar $\chi$
is 
\begin{equation}
 <\chi>_{s} \equiv [\frac {\int 
\chi \sqrt{^3|g|}d^3x} {\int
 \sqrt{^3|g|}d^3x }]_{\rm{limit~over~entire~space}}
\end{equation}
$<\chi>_s$ is thus invariant for all transformations involving the space coordinate $x^i$
only. 

We can orient the coordinates such that $\dot v^\alpha$ has only one nonvanishing 
component say along the coordinate $x^1$. As $\dot v^\alpha v_\alpha = 0, x^1$ is a spacelike coordinate.
Again since with (2), 
\begin{equation}
\dot v_i = \frac {1}{2}[ln (g_{00})]_{,i} , \dot v_0 = 0
\end{equation}
the three space vector $\dot v_i$ is a gradient vector and hence 
hypersurface orthogonal. Hence with the above stipulation

\begin{equation}
 ds^2 = g_{00} dx^{{0}^2} + g_{11}dx^{1^2} + g_{ab}dx^adx^b   
\end{equation}
where $a,b$ run over the indices 2 and 3. One may wonder whether the
metric forms (2) and (5) are globally valid with a single coordinate system.
However, we note that all known regular solutions with non-vanishing
$T_{\mu \nu}$ admit a single coordinate system if there be no
discontinuity in $T_{\mu\nu}$. Such discontinuities, although
not inconsistent with the condition of regularity, seems unappealing 
in a cosmological model and in our discussion we shall assume that the 
forms (2) and (5) are valid over entire space time with a single
coordinate system. Obviously $g_{00}$ is a function of $x^1$
and maybe $x^0$ and $g_{11}, g_{ab}$ may be functions of all
the four coordinates.
As the tangent vector to $x^1$ coordinate line is a gradient, $x^1$
lines cannot form a closed loop. They may either run from $-\infty $ to $+\infty$
or in case they diverge from a point (as the radial lines in case
of spherical or axial symmetry) they may run from zero to infinity. 
In any case if 
$\int^\infty \sqrt |g_{11}| dx^1$ 
converges to a finite value 
(i.e., $g_{11} \rightarrow 0$ as $x^1 \rightarrow +\infty$),
then if there be no singularity at infinity one can see by a transformation that the 
$x^1$ lines are closed. (cf. the closed Friedmann universe in which
$\int_{0}^\infty \frac {dr}{(1 + r^2/4)}$ converges
and one can transform $r$ to an angular coordinate $\chi$ with
domain $0$ to $2\pi$.) Again this would make the gradient vector vanish everywhere. We have thus
a nontrivial $\dot v ^\alpha $ only if 
$\int^\infty \sqrt |g_{11}| dx^1$ 
diverges.

Again,  the scalar $\dot v^{\alpha} _{;\alpha}$ must vanish or oscillate
about a mean vanishing value as $x^1 \rightarrow \pm \infty$ as otherwise
the norm of $\dot v^\alpha$ would blow up - this is apparent when 
one recalls that in absence of singularity, the covariant divergence reduces
to the ordinary divergence 
in a locally Lorentzian coordinate system. Now the space average of $\dot v^\alpha _{;\alpha}$ is
\begin{equation}
 <\dot v^{\alpha}_{;\alpha}>_{s} 
\equiv [\frac {\int 
 \dot v^{\alpha}_{;\alpha} 
 \sqrt{^3|g|}d^3x }
 {\int \sqrt{^3|g|}d^3x }]_{\rm{limit~over~entire~space}}
\end{equation}
 
If $\sqrt{^3|g|}$ diverges or remains finite at infinity, the denominator 
integral diverges and the vanishing of $\dot v^{\alpha}_{;\alpha}$
(or its mean value) at infinity will make the divergence of the numerator
integral weaker. Consequently in the limit 
$ <\dot v^{\alpha}_{;\alpha}>_s $ would vanish. In case 
$ \sqrt{^3|g|}$ vanishes at infinity, this will be due to the vanishing 
of the two dimensional determinant $|g_{ab}|$ as we have seen that for nontrivial 
$\dot v^\alpha$, $g_{11}$ cannot vanish. Thus, in this case, as this factor
is common to both the denominator and the numerator integrals, the vanishing
of 
 $\dot v^{\alpha}_{;\alpha} $
at infinity again ensures  
 $<\dot v^{\alpha}_{;\alpha}>_s = 0 $

Note that in eq (1), all the terms on the left are positive
definite as we are assuming the strong energy condition. Hence 
with 
 $<\dot v^{\alpha}_{;\alpha}>_s = 0 $,
it follows,
\begin{equation}
-<\dot \theta>_s \ge \frac {1} {3} <\theta^2>_s
\end{equation}

It may happen  that $\dot \theta$ and $\theta^2$ both vanish at spatial 
infinity such that the relation (7) is an equality with both
sides vanishing. That will lead to the result that space average of all
the scalars in (1) vanish and thus prove our theorem. If that is not so, 
then either at every point  
\begin{equation}
-\dot \theta \ge \frac {1} {3} \theta^2
\end{equation}
which will lead to a blow up of $\theta$ in the finite past or future
or that in some regions of each space section
\begin{equation}
-\dot \theta  > \frac {1} {3} \theta^2
\end{equation}

Integrating over the  $x^0$ lines one finds
a blowing up of $\theta$ in the finite past or future. As already noted,
$\theta$ is a scalar formed from the Ricci tensor components, and so it cannot blow
up in a nonsingular solution.  Hence we conclude 
that (7) must
be an equality with both sides vanishing and thus our theorem is proved. 

In particular we note that if the space is closed so that the total
spatial volume is finite, the theorem implies that the positive definite
scalars in (1) will vanish everywhere or in other words there is no nontrivial
singularity free solutions in case of closed space sections. 

It may be worthwhile to make a comparison of the present
theorem with that of Hawking and Penrose. As we have already
remarked the Hawking-Penrose  theorem is of little relevance so far as realistic models
of the universe are concerned as closed trapped surfaces seem inevitable. On the other
hand the present theorem depends on the consideration 
of infinite space integrals and hence it may overlook localized
singularities which do not affect the infinite integrals. Such singularities
are apparently taken care of by the trapped surface condition in Hawking Penrose
theorem.

The author's thanks are due to the members of the
Relativity and Cosmology Centre, Jadavpur University and to
Prof Naresh  Dadhich of IUCAA, Pune for interesting discussions. 

\end{multicols}

\end{document}